\documentclass[final,3p,times,twocolumn]{elsarticle}
\usepackage{amssymb,amsmath}
\usepackage{graphicx}
\usepackage[pdftex,colorlinks=true]{hyperref}

\journal{Physics Letters B}

\begin{document}

\begin{frontmatter}
  \title{Constraining  holographic technicolor}

\author[inr]{D.G. Levkov}\ead{levkov@ms2.inr.ac.ru}
\author[inr,msu]{V.A. Rubakov}
\author[inr]{S.V. Troitsky}
\author[inr]{Y.A. Zenkevich}
\address[inr]{Institute for Nuclear Research of the Russian Academy of
    Sciences, 60th October Anniversary Prospect 7a, Moscow 117312, Russia}
\address[msu]{Physics Department, Moscow State University, Vorobjevy Gory,
Moscow 119991, Russia}
\begin{abstract}
    We obtain a new bound on the value of Peskin-Takeuchi $S$
    parameter in a wide class of bottom-up holographic models for
    technicolor. Namely, we show that weakly coupled holographic
    description in these models implies $S\gg 0.2$. Our bound
    is in conflict with the results of electroweak precision
    measurements, so it strongly disfavors the models we consider.
\end{abstract}

  \begin{keyword}
    Electroweak symmetry breaking, Holographic duality, Oblique
    parameters, Tree unitarity.
  \end{keyword}
\end{frontmatter}

%%%%%%%%%%%%%%%%%%%%%%%%%%%%%%%%%%%%%%%%%%%%%%%%%%%%%%%%%%%%%%%%%%%%%%%%%
\section{Introduction}
\label{sec:introduction}
Long--anticipated discovery~\cite{higgs} of a Higgs--like particle
  brings us face-to-face with major challenge of comprehending in
  detail the mechanism of electroweak symmetry
  breaking~\cite{think}. An interesting option potentially leading to
  composite Higgs is provided by strongly interacting models similar to 
technicolor~\cite{technicolor}.
The latter is usually represented by a gauge  
theory with chiral symmetry $\mbox{SU}(N_f)_L
\times \mbox{SU}(N_f)_R$ which breaks down to $\mbox{SU}(N_f)_V$ in
QCD-like manner; $N_f$ is the number of techniflavors. Electroweak
symmetry is a gauged $\mbox{SU}(2)_L\times \mbox{U}(1)_Y$ subgroup of
the chiral group, so it is broken due to chiral symmetry breaking. 
Unfortunately, the simplest, literally drawn from QCD
technicolor models were ruled out long ago, as they predict
unacceptably large values of the
Peskin-Takeuchi~\cite{peskin} $S$ parameter. This
leaves open~\cite{S_walking} a ``walking''
version~\cite{walking_technicolor} which, however, lacks contact with
the phenomenological information
accumulated by hadron physics.

The gauge/gravity holographic duality~\cite{AdS_CFT_classics} enters
at this stage as an approach to studying technicolor models in terms
of their weakly coupled gravity duals in five 
dimensions~\cite{hong, hirn2, hirn, agashe, piai, top_down}. 
Using the holographic
dictionary~\cite{AdS_CFT_classics}, one  
relates conserved currents $j_\mu^L$, $j_\mu^R$ of the left and right
$\mbox{SU}(N_f)$ chiral groups to five-dimensional gauge
fields\footnote{Hereafter $\mu,\nu=0\dots 3$ and $M,N=0\dots3, 5$.}
$L_M$ and $R_M$. This promotes the global $\mbox{SU}(N_f)_L \times
\mbox{SU}(N_f)_R$ symmetry of the original model to the gauge symmetry
of the holographic dual. The problem of computing current correlators
then reduces to that of solving
classical equations for the dual fields.

Since dual descriptions of realistic technicolor theories are
unknown (see, however, Refs.~\cite{QCD_top_down,top_down}), one
tends to adopt a bottom-up  approach~\cite{bottom-up}
trying to guess the field content and Lagrangian of the
five-dimensional dual model on phenomenological grounds. To this end
one introduces new fields besides $L_M$ and $R_M$, for instance,
an $\mbox{SU}(N_f)_L\times \mbox{SU}(N_f)_R$ bifundamental scalar $X$
representing techniquark condensate~\cite{hong, hirn, agashe,
  piai}. One also selects appropriate conditions at the boundaries of 
the 5D space and allows for departures from the $\mbox{AdS}_5$
geometry~\cite{hirn2,hirn,agashe}. The price to pay is the absence of
an ultraviolet completion of the model which therefore has the status
of an effective theory below a certain UV cutoff.

This simple picture is far from being justified in any
  rigorous sense. Nevertheless, one hopes
  that such models capture essential features of strongly
  coupled dynamics and therefore serve as good toy models for
  technicolor theories.

In this Letter we derive a new constraint on a class of
holographic technicolor models, namely, those~\cite{hirn, agashe}
containing two ${\mbox{SU}(N_f)}$ gauge fields $L_M$, $R_M$ and
bifundamental $X$. The fields live in an interval in the warped fifth
dimension, with boundary conditions to be specified below. 
  We show that weakly coupled description of these holographic models
  implies large values of $S$ parameter. Namely,
${S\gg 0.2}$, otherwise: (i) the UV cutoff drops below $6\pi m_W/g \sim
2.5\,\mbox{TeV}$; (ii) 
correlators of electroweak currents with momenta exceeding the UV
cutoff are sensitive to strongly coupled sector of the 5D theory and
therefore not tractable; (iii) no reliable predictions for the
spectrum can be made. Properties (i)---(iii) degrade the status of the
holographic technicolor models 
to that of theories with massive $W$ bosons and no Higgs
mechanism: the latter are also strongly coupled above a few
$\mbox{TeV}$. On the other hand, the constraint $S\gg 0.2$ is in
conflict with the experimental result
~\cite{PDG} $S=-0.07\pm 0.1$ and therefore
strongly disfavors the models.

We introduce the models in Sec.~\ref{sec:model},  review their
spectrum and computation of $S$ in
Secs.~\ref{sec:vector-spectrum} and~\ref{sec:s-parameter}, respectively. In
Sec.~\ref{sec:strong-coupling} we present a derivation of the weak
coupling condition in general warped background. On this basis we
obtain new bound on 
$S$ in
Sec.~\ref{sec:constr-s-param}.
In Sec.~\ref{sec:high-order-oper} we show that our bound
 is stable with respect to 
the addition of higher-order operators to the Lagrangian. 
We summarize in Sec.~\ref{sec:conclusions-1}.

%%%%%%%%%%%%%%%%%%%%%%%%%%%%%%%%%%%%%%%%%%%%%%%%%%%%%%%%%%%%%%%%%%%%%%
\section{Models}
\label{sec:model}
The models we consider~\cite{hirn, agashe} are formulated in a patch of
5D space with warp factor $w(z)$,
$$
ds^2 = w^2(z) \left( \eta_{\mu\nu}dx^\mu dx^\nu - dz^2 \right)\;,
\qquad z\in 
    [z_{\mathrm{UV}},\, z_{\mathrm{IR}}]\;,
$$
where  $w(z_{\mathrm{UV}})=1$. The
action reads,
\begin{multline}
\label{eq:1}
{\cal S} = \int dz\, d^4 x   \, \mathrm{tr}\, \Big[ w(z)
  \left(L_{MN}^2 + R_{MN}^2\right)/2g_5^2 \\
  +  w^3(z) D_M X^\dag D^M X - w^5(z) V(X) \Big] \; .
\end{multline}
It
describes two $\mbox{SU}(N_f)$ gauge fields $L_M$ and $R_M$
interacting with scalar $X$; $g_5$ is the five-dimensional
gauge coupling. Hereafter the integrals over $z$ 
run from $z_{\mathrm{UV}}$ to $z_{\mathrm{IR}}$; we write
$w(z)$ explicitly  and convolve indices with mostly negative flat
metric. In our notations $L_M$ and $R_M$ are
anti-Hermitean matrices, $L_{MN} = \partial_{[M}
  L_{N]} + L_{[M} L_{N]}$. The bifundamental scalar $X$ is gauge
transformed as $X\to \omega_L X\omega^\dag_R$, where $\omega_{L,R} \in
\mbox{SU}(N_f)_{L,R}$; its covariant derivative is  $D_M X =
  \partial_M X + L_M X - XR_M$.

We assume that the models (\ref{eq:1}) are dual to
strongly coupled technicolor theories. Then
$\mbox{SU}(N_f)_L \times \mbox{SU}(N_f)_R$ gauge symmetry must be
broken to the diagonal subgroup $\mbox{SU}(N_f)_V$.  To achieve this, we
invoke
two sources of symmetry breaking  that work together~\cite{hirn,
  agashe}.
One is the boundary
conditions at the IR brane, 
\begin{equation}
\label{eq:4}
L_\mu = R_\mu\;, \qquad \partial_z L_\mu = - \partial_z R_\mu \qquad
\mbox{at} \;\; z = z_{\mathrm{IR}}\;,
\end{equation}
and another is
the vacuum profile of $X$ which is assumed\footnote{The scalar
  potential $V(X)$ and boundary conditions for $X$ should be chosen
  accordingly.} to have the form ${X_0 =v(z)\cdot\mathbb{I}}$, where
$\mathbb{I}$ is the $N_f\times N_f$ unit matrix, $v(z)$ is real. 
The conditions~(\ref{eq:4}) and
vacuum $X_0$ are preserved by the diagonal gauge transformations
with $\omega_L = \omega_R$ and
$\partial_z\omega_L|_{z_{\mathrm{IR}}}=\partial_z
  \omega_R|_{z_{\mathrm{IR}}}=0$, 
so the diagonal subgroup $\mbox{SU}(N_f)_V$ remains unbroken.

We do not consider theories~\cite{hirn1,hirn2} with explicit breaking 
of gauge invariance in the bulk and accidentally enlarged gauge symmetry at the quadratic level\footnote{One can 
formally restore gauge invariance  
by introducing St\"uckelberg/spurion fields~\cite{hirn1,hirn2}. This 
  does not make a theory healthy.}, as these properties  generically
lead to pathologies: strong coupling at all scales, ghosts, etc.

The models we consider are parametrized by the coupling constant
$g_5$, warp factor $w(z)$, and vacuum profile $v(z)$. We note
that a subclass of models without the scalar $X$~\cite{Higgsless} is
effectively obtained at $v(z)=0$; gauge symmetry in this case is
broken by the boundary conditions (\ref{eq:4}). Our analysis applies
at $v(z)=0$ equally well. We impose
consistency requirements: (i) $v^2(z) \ll \Lambda_5^3$, where
$\Lambda_5$ is a UV cutoff of the models~(\ref{eq:1}); (ii) 
$w(z)$ and $v(z)$ do not vary on the 
  physical length scale of order $w(z) \Delta z \sim \Lambda_5^{-1}$.
The conditions (i), (ii) ensure the suppression of higher--order operators 
$(X^\dag X)^n /\Lambda_5^{3n}$, $(D_M X D_M 
    X^\dag)^n/\Lambda_5^{5n}$, etc., which are present in 
the general effective Lagrangian.

By construction, the fields $L_M$ and $R_M$ are dual to the chiral
currents $j_\mu^L$, $j_\mu^R$ of the technicolor theory. This
means~\cite{AdS_CFT_classics} that the current correlators are
computed holographically in terms of $L_M$ and $R_M$. First, one solves
the classical field equations with the boundary conditions (\ref{eq:4}) and
\begin{equation}
\label{eq:2}
L_\mu\big|_{z_{\mathrm{UV}}} = \bar{L}_\mu(x)\;, \qquad
R_\mu\big|_{z_{\mathrm{UV}}} = \bar{R}_\mu(x)\;.
\end{equation}
Second, one computes the action~(\ref{eq:1}) for the solution,
to obtain the functional ${\cal S}={\cal S}[\bar{L},\, \bar{R}]$. In the
holographic approach ${\cal S}$ is interpreted as a  generating
functional~\cite{AdS_CFT_classics,bottom-up}
for correlators of the chiral currents. In
particular,
\begin{equation}
\label{eq:3}
\langle j_\mu^{La}(x) j_\nu^{Rb}(y) \rangle = -i \left.\frac{\delta^2
  {\cal S}}{\delta \bar{L}^{\mu a}(x) \delta \bar{R}^{\nu b}(y)}
\right|_{\bar{L} = \bar{R} = 0}\;,
\end{equation}
where the component fields
$L_\mu^a = 2i \, \mathrm{tr}(L_\mu t^a)$ and $R_\mu^a$ are
introduced. The field content of the four-dimensional
technicolor theory remains unknown in the bottom-up holographic
approach: the theory is defined by correlators like~(\ref{eq:3}).

To add electroweak interactions, we consider the 4D picture and embed
 exactly one\footnote{In other models~\cite{technicolor} one
    embeds electroweak group $N_f/2$ times and obtains $N_f/2$ times
    larger value of $S$ parameter.} copy of $\mbox{SU}(2)_L$ and
$\mbox{U}(1)_Y$  
electroweak groups into the left 
and right $\mbox{SU}(N_f)$ chiral groups. We couple
the respective isospin components $j_{\mu}^{L\bar{a}}$ and
$j_\mu^{R3}$ of the chiral currents to the $\mbox{SU}(2)_L$
and $\mbox{U}(1)_Y$ electroweak bosons, where $\bar{a}=1\dots 3$. This
corresponds to gauging $\mbox{SU}(2)_L\times 
\mbox{U}(1)_Y$ subgroup of the global flavor group. The electroweak
symmetry is then spontaneously broken due to chiral symmetry
breaking. We invoke 5D description by noting that  electroweak
observables are related to the current correlators which, in turn,
are computed via Eq.~(\ref{eq:3}). For example, the polarization
operator between the $\mbox{SU}(2)_L$ gauge field and hypercharge
field is equal to
\unitlength 1mm
\begin{align}
\begin{minipage}{20mm}
\begin{picture}(20,8)
\put(0,1){\includegraphics[width=2cm]{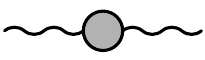}}
\put(1.5,6){$W_\mu^{\bar{a}}$}
\put(15,6){$B_\nu$}
\put(7.5,-0.3){$\longrightarrow$}
\put(9,-1.5){$p$}
\end{picture}
\end{minipage}
& = i \eta_{\mu\nu} g g' \Pi_{\bar{a}Y} (p^2 ) + p_\mu p_\nu \mbox{-
  terms}\label{eq:7}\\
& = - gg' \int d^4 x\, \mathrm{e}^{ipx} \langle j_{\mu}^{L{\bar{a}}}(x)
j_\nu^{R3}(0) \rangle \notag\;,
\end{align}
where $g$ and $g'$ are the
electroweak gauge couplings.

%%%%%%%%%%%%%%%%%%%%%%%%%%%%%%%%%%%%%%%%%%%%%%%%%%%%%%%%%%%%%%%%%%%%%
\section{Spectrum}
\label{sec:vector-spectrum}
In quest for constraining the models~(\ref{eq:1}) we need information
about their spectra. Since the extent of the fifth coordinate is finite,
there is a discrete tower of Kaluza-Klein modes which are 
interpreted as technimesons. Below we analyze vector excitations
and leave aside scalars\footnote{Including the Nambu-Goldstone
  bosons (technipions)~\cite{bottom-up}.}
whose spectrum  depends on the form of the potential $V(X)$.

One notices that the action (\ref{eq:1}), boundary conditions
(\ref{eq:4}) and vacuum profile $X_0$
are invariant under $\mathbb{Z}_2$ parity transformations
$L_M \leftrightarrow R_M$, $X \leftrightarrow X^\dag$. Thus, linearized
equations for the parity-even vector field $V_M = (L_M +
R_M)/\sqrt{2}$ decouple from equations for the parity-odd axial-vector
field $A_M = (L_M -R_M)/\sqrt{2}$. In the $V_5 = A_5 = 0$ gauge, it is consistent
to set
$\partial_\mu V^\mu= \partial_\mu A^\mu = 0$, and the field
equations become
\begin{subequations}
\label{eq:16}
\begin{align}
&-\frac{1}{w}\partial_z \left(w \partial_z V_\mu \right) - p^2 V_\mu =
  0\;,\label{eq:8}\\
&-\frac{1}{w}\partial_z \left(w \partial_z A_\mu \right) - (p^2 -
  2g_5^2w^2 v^2) A_\mu = 0\label{eq:9}\;,
\end{align}
\end{subequations}
where $p_\mu$ is 4D momentum.
Since $V_\mu$ is the gauge field of the unbroken diagonal
subgroup, symmetry-breaking effects due to $v(z)\ne 0$ are felt only
by $A_\mu$. We supplement Eqs.~(\ref{eq:16}) with boundary conditions
\begin{equation}
\label{eq:10}
V_\mu\big|_{z_{\mathrm{UV}}} = \partial_z V_\mu\big|_{z_{\mathrm{IR}}}
=0\;, \qquad
 A_\mu\big|_{z_{\mathrm{UV}}} = A_\mu\big|_{z_{\mathrm{IR}}}=0\;,
\end{equation}
deduced from Eqs.~(\ref{eq:4}) and~(\ref{eq:2}). 
Equations~(\ref{eq:16}), (\ref{eq:10}) form two independent boundary
value problems for the vector and axial-vector mass spectra $p^2 =
(m_n^V)^2$ and ${p^2 = (m_n^A)^2}$; we denote the respective
eigenfunctions by $V_n(z)$ and $A_n(z)$. The normalization
condition follows from (\ref{eq:1}), it reads: ${\int
dz\, w(z) V_n(z) V_{n'}(z) = \delta_{nn'}}$ and likewise for $A_n(z)$.

It is not possible to find the spectra for arbitrary $w(z)$
and $v(z)$. There are some general properties, however.
First, the operators in Eqs.~(\ref{eq:16}) and hence eigenvalues
$(m_n^V)^2$, $(m_n^A)^2$ are positive-definite.
Second, the axial-vector masses are larger\footnote{At $v(z)=0$ this
  is the consequence of the fact that $A_n(z)$ satisfy the same equation
  as $V_n(z)$, but with the Dirichlet boundary condition at
  $z=z_{\mathrm{IR}}$
  instead of the Neumann one. At $v(z)\ne 0$ the axial masses are shifted
  further upwards because the additional term in Eq.~(\ref{eq:9}) is
  positive.}, $m_n^A \geq m_n^V$.

One learns more from the vector Green's function
\begin{equation}
\label{eq:5}
G_p^V(z,z') = -\sum_{n} \frac{V_n(z) V_n(z')}{p^2 - (m_n^V)^2}\;,
\end{equation}
which satisfies the boundary conditions~(\ref{eq:10}) for vectors and
Eq.~(\ref{eq:8}) with $\delta(z-z')/w(z)$ in the right-hand side.
One solves these equations at $p^2 = 0$,
\begin{equation}
\label{eq:6}
G_{p=0}^V (z,z') = \theta(z-z') I(z') + (z\leftrightarrow z')\;,
\end{equation}
where $I(z) = \int_{z_{\mathrm{UV}}}^z dz'/w(z')$. Combining
Eqs.~(\ref{eq:5}) and  (\ref{eq:6}), one finds a sum rule for the vector
masses~\cite{hirn},
$$
\int dz \, w(z) \,G_{p=0}^V(z,z) =
\sum_{n} \frac{1}{(m_n^V)^2} = \int dz\,
w(z) I(z)\;.
$$
This relation sets a bound on the mass $m_1^V$ of the lightest vector
technimeson:
\begin{equation}
\label{eq:14}
\frac{1}{(m_1^V)^2} \leq \int dz\,
w(z) I(z)\; .
\end{equation}
Since ${m_n^A \geq
m_n^V}$, the axial-vector masses are also bounded by the right-hand side of
Eq.~(\ref{eq:14}).

The axial-vector Green's function $G_p^A(z,z')$ is defined in a similar way,
as a solution to Eq.~(\ref{eq:9}) with ${\delta(z-z')/w(z)}$ in the
right-hand side and boundary conditions~(\ref{eq:10}) for $A_{\mu }$.
At $p^2=0$ it can be expressed via a
particular solution $a(z)$ of Eq.~(\ref{eq:9}) satisfying
${a(z_{\mathrm{UV}}) = 1}$, $a(z_{\mathrm{IR}}) = 0$. One obtains,
\begin{equation}
\label{eq:15}
G_{p=0}^A(z,z') = \theta(z-z') \, a(z)a(z')I_A(z')+ (z\leftrightarrow
z')\;,
\end{equation}
where $I_A(z) = \int_{z_{\mathrm{UV}}}^z dz'/[w(z')a^2(z')]$. 

%%%%%%%%%%%%%%%%%%%%%%%%%%%%%%%%%%%%%%%%%%%%%%%%%%%%%%%%%%%%%%%%%%%%%%
\section{$S$ parameter}
\label{sec:s-parameter}
The Peskin-Takeuchi $S$ parameter~\cite{peskin} measures contributions of
new physics to the polarization operator $\Pi_{3Y}$,
\begin{equation}
\label{eq:20}
S = -16 \pi\,\frac{d\Pi_{3Y}}{dp^2}\Bigg|_{p^2=0}\; .
\end{equation}
The value of $S$ is
extracted from the electroweak precision measurements.

We evaluate $S$ by the holographic recipe~(\ref{eq:3}), (\ref{eq:7}). 
Equation~(\ref{eq:3}) involves only quadratic part of the
action, so we solve linear equations~(\ref{eq:16}) with boundary
conditions~(\ref{eq:4}), (\ref{eq:2}),
\begin{align}
\notag
&V_\mu(p,z) = \bar{V}_\mu(p) + p^2\bar{V}_\mu(p)
\int dz'\, w(z')G_p^V(z,z')\,,\\
\label{eq:27}
&A_\mu(p,z) = \bar{A}_\mu(p)a(z) \\ &\notag \qquad
\qquad+ p^2\bar{A}_\mu(p)
\int dz'\, w(z') \,G_p^A(z,z') \, a(z')\;,
\end{align}
where $a(z)$ is 
defined in the previous section,
$\bar{V}$ and $\bar{A}$ are the linear combinations of $\bar{L}$ and
$\bar{R}$. Upon integrating by parts, one writes for  the quadratic
part of the action 
\begin{equation}
\notag
{\cal S}^{(2)} 
  = \frac1{g_5^2}\left.\int d^4x \,\mathrm{tr}\,(V_\mu \partial_z
V_\mu + A_\mu \partial_z A_\mu)\right|_{z_{\mathrm{UV}}}\; .
\end{equation}
We substitute
solutions (\ref{eq:27}) into the action and vary it with
respect to $\bar{L}$, $\bar{R}$. The result for $\Pi_{3Y}$ 
is
\begin{align}
\label{eq:12}
 &\Pi_{3Y}(p^2) = \frac{1}{2g_5^2} \partial_z a \Big|_{z=z_{\mathrm{UV}}}
\\\notag
 &\;\;\; - \frac{p^2}{2g_5^2}\int dz' w(z') \partial_z
\left( G_p^V(z,z') - G_p^A(z,z')a(z') \right)
\Big|_{z=z_{\mathrm{UV}}}.
\end{align}
We finally compute $S$ parameter~\cite{hirn2,hirn,agashe}: 
\begin{equation}
\label{eq:26}
S = \frac{8\pi}{g_5^2} \int dz \, w(z) \left[1 -
  a^2(z)\right]\;,
\end{equation}
where the explicit Green's functions (\ref{eq:6}), (\ref{eq:15}) at
${p^2=0}$ were used. We remind that $a(z)$ satisfies Eq.~(\ref{eq:9})
with $p^2=0$ and boundary conditions $a(z_{\mathrm{UV}}) = 1$,
${a(z_{\mathrm{IR}}) = 0}$.

The first term in Eq.~(\ref{eq:12}) does not depend on $p^2$ and
therefore represents the $Z$-boson mass:
\unitlength 1mm
\begin{equation}
\nonumber
i\eta_{\mu\nu} gg' \Pi_{3Y}(0) =
\begin{minipage}{16mm}
\begin{picture}(16,6)
\put(1,0){\includegraphics[width=1.5cm]{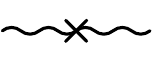}}
\put(0,5.5){$W_\mu^3$}
\put(13,5.5){$B_\nu$}
\put(7,5.5){$m_Z^2$}
\end{picture}
\end{minipage}
 = -i \eta_{\mu\nu}\cos \theta_W \sin\theta_W m_Z^2\;.
\end{equation}
Here we ignored $p_\mu p_\nu$-terms and introduced the weak mixing angle,
$\mathrm{tg}\,\theta_W = g'/g$. The expression~(\ref{eq:12}) gives
\begin{equation}
\label{eq:13}
m_W^2 = m_Z^2 \cos^2\theta_W = -\frac{g^2}{2g_5^2} \,\partial_z
a\big|_{z_{\mathrm{UV}}}\; ,
\end{equation}
where the first equality is a consequence of the custodial symmetry inherent
in the models~(\ref{eq:1}).
Non-zero masses of $W$ and $Z$ bosons are manifestations of the electroweak symmetry
breaking, cf. Refs.~\cite{bottom-up, LRTZ2}.

In Ref.~\cite{agashe} it was proven that $S>0$  in
the class of models we consider.
This is seen from Eq.~(\ref{eq:26}): 
the function $f(z) =
aw\partial_z a$ is negative, since $\partial_z f>0$
and $f(z_{\mathrm{IR}}) =0$ due to Eq.~(\ref{eq:9}) and
$a(z_{\mathrm{IR}}) =0$. In other words, $\partial_z a^2
< 0$, i.e. $a^2(z)$ monotonically decreases from
$a^2(z_{\mathrm{UV}})=1$ to $a^2(z_{\mathrm{IR}})=0$ implying
${a^2<1}$ and $S>0$.

Below we further constrain the value of $S$ by making use of an additional
requirement of weak coupling.

%%%%%%%%%%%%%%%%%%%%%%%%%%%%%%%%%%%%%%%%%%%%%%%%%%%%%%%%%%%%%%%%%%%%%%
\section{Weak coupling condition}
\label{sec:strong-coupling}
The model (\ref{eq:1}) is non-renormalizable and therefore makes sense
below some energy cutoff $\Lambda_5$. In flat spacetime
$\Lambda_5$ is computed from the partial amplitudes for gauge boson
scattering. On dimensional grounds these are proportional to $g_5^2
P$, where  $P$ is a 5D momentum. The amplitudes grow with energy and
break unitarity bound at ${P\gtrsim 1/g_5^2}$ signaling strong
coupling. Thus, ${\Lambda_5\sim 1/g_5^2}$.

In warped spacetime the situation is more
subtle~\cite{strong,sundrum}. Correlators from the UV brane to UV brane,
such as~(\ref{eq:3}), are functions of the  conformal momentum $p$.
On the other hand, scattering at ${z=z_0}$ is perturbative if the
local physical 
momentum $P=p/w(z_0)$ satisfies $P\ll \Lambda_5$. Thus,
brane-to-brane correlators are completely in the weak coupling regime
at ${p \ll \Lambda_5 w_{\min}}$, where $w_{\min}$ is the minimal value
of $w(z)$. They can still be tractable at higher momenta if
contributions from the strongly coupled region $w(z) < p/\Lambda_5$
are suppressed.

Let us compute the UV cutoff for the general background $w(z)$. This
generalizes the analysis of Refs.~\cite{Higgsless} performed in the
case of flat metric. To this 
end we consider the  
amplitude ${\cal A}_{nn'\to mm'}$ for the vector-mode scattering
${V_n^aV_{n'}^b \to V_{m}^a V_{m'}^b}$. At the tree level, this amplitude
is the sum of a $V^4$ vertex
(\begin{minipage}{3mm}
\includegraphics[width=3mm]{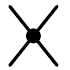}
\end{minipage})
and exchange diagrams
(\begin{minipage}{4.2mm}
\includegraphics[width=4.2mm]{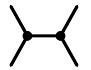}
\end{minipage}).
The vertices $VVA$ and $VVX$ are forbidden by parity conservation and
$\mbox{SU}(2)_V$ gauge symmetry, respectively. What remains are
the diagrams involving $V_\mu$ only,
\begin{equation}
\nonumber
\begin{minipage}{20mm}
\unitlength 1mm
\begin{picture}(20,12)
\put(4,0){\includegraphics[width=12mm]{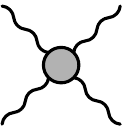}}
\put(0,0.5){$V_n^a$}
\put(0,9.2){$V_{n'}^b$}
\put(16,0.5){$V_m^a$}
\put(16,9.2){$V_{m'}^b$}
\end{picture}
\end{minipage}
=
\begin{minipage}{11mm}
\unitlength 1mm
\begin{picture}(11,7)
  \put(0,0){\includegraphics[width=11mm]{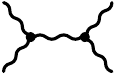}}
  \put(5,7){$s$}
\end{picture}
\end{minipage}
+
\begin{minipage}{11mm}
\unitlength 1mm
\begin{picture}(11,7)
  \put(0,0){\includegraphics[width=11mm]{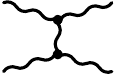}}
  \put(5,7){$t$}
\end{picture}
\end{minipage}
+
\begin{minipage}{11mm}
\unitlength 1mm
\begin{picture}(11,7)
 \put(0,0){\includegraphics[width=11mm]{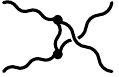}}
  \put(5,7){$u$}
\end{picture}
\end{minipage}
+
\begin{minipage}{11mm}
\unitlength 1mm
\begin{picture}(11,7)
  \put(0,0){\includegraphics[width=11mm]{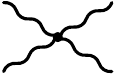}}
\end{picture}
\end{minipage}
\end{equation}
We calculate the amplitude at high energies when
many Kaluza-Klein modes are ultrarelativistic --- as well as the
colliding particles. For the latter, we consider longitudinal polarizations
${\epsilon^\mu(p) \approx   p^\mu/m}$ and isospin states
$(ab)\to (ab)$. We obtain\footnote{Calculations
  simplify in the ${\cal R}_\xi$ gauge~\cite{R_xi} where
  the longitudinal components of massive vector modes can be traded  at
  high energies for  Nambu-Goldstone bosons.},
\begin{equation}
\label{eq:17}
{\cal A}_{nn'\to mm'} = -g_5^2\, d^{ab}\,  g_{nn'mm'} \, \frac{3 +
    \cos^2\theta}{2+2\cos\theta}\,.
\end{equation}
Here $\theta$ is the scattering angle, $d^{ab} = \sum_c
  (f^{abc})^2$ involves the $\mbox{SU}(N_f)$ structure constants
  $f^{abc}$, $g_{nn'mm'} = \int dz \,w \phi_n \phi_{n'} \phi_m
    \phi_{m'}$ is the overlap integral of functions ${\phi_n =
  \partial_z V_n / m_n^V}$. To understand the meaning of $\phi_n$, one
performs the 
gauge transformation which eliminates
longitudinal components $V_\mu^L = ip_\mu V^L$ and induces instead
  $V_5 = \partial_z  V^L$. One sees that $\phi_n $ are the 
  wave functions of the longitudinal modes; they satisfy completeness
  relation $\sum_n \phi_n(z)\phi_n(z') = \partial_z \partial_{z'}
  G_{p=0}^V(z,z') = \delta(z-z')/w(z)$, where Eqs.~(\ref{eq:5}),
  (\ref{eq:6}) were used.

We expect that in terms of conformal momentum, the cutoff depends on
$z$. To see this explicitly, we localize colliding particles in the
fifth dimension by considering the Kaluza-Klein state $|V_{z_0}\rangle
= {\cal N}\sum_{n<n_0}\phi_n(z_0) |V_n\rangle$, where ${\cal N}$ is a
normalization constant. At $n_0\gg 1$, the wave function of this state
is concentrated near $z= z_0$, as the completeness of $\phi_n$ suggests.
Such a localization is consistent with the presence of the UV cutoff, since,
as we pointed out in Sec.~\ref{sec:model}, the function $w(z)$ does not
strongly vary on the physical distance scale $\Lambda_5^{-1}$.
The amplitude of the process $V_{z_0} V_{z_0} \to V_{z_0} V_{z_0}$ is
\begin{equation}
\label{eq:21}
{\cal A}_{z_0} = {\cal N}^4 \sum_{nn'mm'<n_0}
\phi_n^{(z_0)} \phi_{n'}^{(z_0)} \phi_m^{(z_0)} \phi_{m'}^{(z_0)}
    {\cal A}_{nn' \to mm'},
\end{equation}
where $\phi_n^{(z_0)} = \phi_n(z_0)$.

Let us now recall the unitarity conditions $|\mathrm{Re}\, {\cal A}_l|
\leq 1/2$ for partial amplitudes, where $l$ is the
angular momentum. Particularly useful is the constraint
$$
|\mathrm{Re}\,({\cal A}_0 + {\cal A}_1)| \equiv \frac{1}{32\pi}\left|
\int_{-1}^1 d\mathrm{cos}\theta\, (1 + \cos\theta)\, \mathrm{Re}\,{\cal
  A}\right| \leq 1
$$
where the left-hand side is free of collinear divergences. Making use of
Eqs.~(\ref{eq:17}), (\ref{eq:21}) and explicitly writing 
$g_{nn'mm'}$, we find
\begin{multline}
\label{eq:18}
|\mathrm{Re}\,({\cal A}_0 + {\cal A}_1)|_{z_0} =
\frac{5g_5^2 }{48\pi}\, d^{ab}
{\cal N}^4 
\sum_{nn'mm'<n_0}\phi_n^{(z_0)} \phi_{n'}^{(z_0)} \\ \times\phi_{m}^{(z_0)}
\phi_{m'}^{(z_0)}  \int dz \, w(z)\phi_n^{(z)}
\phi_{n'}^{(z)}\phi_{m}^{(z)}
\phi_{m'}^{(z)} \leq 1 \;.
\end{multline}
We consider the indices $(ab)$, $a\ne b$ belonging to the $\mbox{SU}(2)$
subgroup of $\mbox{SU}(N_f)$ and obtain $d^{ab} = 1$. One sum in
Eq.~(\ref{eq:18}) is  
proportional to $\delta(z-z_0)$ due to completeness of 
$\phi_n$, the others are equal to the semiclassical density
of states $\sum_{n<n_0} \phi_n^2(z_0) \approx \Delta P_{z}/2\pi = m_{n_0}^V
/[\pi w(z_0)]$. The normalization factor of $|V_{z_0}\rangle$ equals
${{\cal N}^2 = \pi w(z_0)/m_{n_0}^V}$. One sees that the inequality
(\ref{eq:18}) takes the form $5g_5^2 m_{n_0}^V \leq 48 \pi^2
w(z_0)$. It bounds the value of the highest available mass $m_{n_0}^V$
and hence  conformal momentum: $p < w(z_0)\Lambda_5$, where
$\Lambda_5 = 48\pi^2/5g_5^2$ is the local scale of strong coupling.

Common sense suggests that theories with too low UV cutoff are
not viable. In the rest of this section we argue
that the model (\ref{eq:1}) is not tractable unless
\begin{equation}
\label{eq:19}
m_1^V \ll \Lambda_5 w_{\mathrm{min}}\;, \qquad \mbox{where} \;\;\;\;
\Lambda_5 = 48\pi^2/5g_5^2 \; .
\end{equation}
Here $m_1^V$ is the lowest vector mass. 

First, one notices that the tower of vector modes is
strongly coupled whenever Eq.~(\ref{eq:19}) is violated. Indeed, all
vector masses are then above the cutoff in the
region ${w(z) <   m_1^V / \Lambda_5}$. Mode amplitudes are large there: a
semiclassical estimate gives $\phi_n^2(z), \, V_n^2(z) \propto 1/w(z)$. Thus,
processes involving vector modes receive large contributions from the
strongly coupled  region ${w(z) < m_1^V / \Lambda_5}$ and cannot be
treated within the effective theory~(\ref{eq:1}). This
prevents one to draw any conclusions about
vector technimesons and hence damages predictability. 

In warped models, one can sometimes consistently consider conformal
momenta exceeding $\Lambda_5 w_{min}$, as long as one deals exclusively with
brane-to-brane correlators~\cite{strong,sundrum}. The point is that
at high Euclidean momenta, the brane-to-bulk propagator decays as 
$\exp[-p (z - z_{UV})]$, which can suppress effects coming from the strongly
coupled region $w(z) < p/\Lambda_5$.
For $p \sim \Lambda_5 w_{min}$ such suppression mechanism requires
$\Lambda_5 w_{min} (z_{IR} - z_{UV}) \gg 1$. This, in turn, 
implies the inequality (\ref{eq:19}), since $m_1^V
(z_{\mathrm{IR}}-z_{\mathrm{UV}}) \simeq \pi/2$ according to
the Bohr-Sommerfeld rule. On the contrary, once the inequality
 (\ref{eq:19}) is violated, $\Lambda_5 w_{min}$ is the true cutoff
for momenta $p$ referring to the UV brane.

Another way to see the strong coupling problem for the
brane-to-brane correlators at ${m_1^V > \Lambda_5
w_{\mathrm{min}}}$ is to consider the propagator in the
form (\ref{eq:5}).
At ${p\lesssim m_1^V}$ it is dominated
by the first term in the sum (\ref{eq:5}) and therefore
proportional to $V_1(z)$. The latter grows with $z$, as the lowest
eigenfunction of Eqs.~(\ref{eq:8}), (\ref{eq:10}). This means that
$G_p^V$ cannot suppress contributions from the
strongly coupled region $w(z) < p/\Lambda_5$ for momenta in the range
$\Lambda_5 w_{min} < p < m^V_1$.

So far we have argued that once the inequality~(\ref{eq:19}) is
violated, the theory makes sense only at $p < \Lambda_5
w_{\mathrm{min}}$. Let us show that the scale
$\Lambda_5w_{\mathrm{min}}$ is unacceptably low, 
even somewhat lower than the  cutoff in a 4D theory of
massive $W$-bosons  without the Higgs mechanism. To this end we use
the Rayleigh-Ritz inequality for the lowest eigenvalue of
Eqs.~(\ref{eq:8}), (\ref{eq:10}),
\[
(m_1^V)^2 \leq \frac{\int
  dz\, w (\partial_z f)^2}{\int dz \, wf^2}\; ,
\]
which holds for arbitrary function $f(z)$ satisfying
$f(z_{\mathrm{UV}})=0$. We select $f(z)=a(z)-1$, where $a(z)$ enters
Eq.~(\ref{eq:13}). Since in the case under consideration $\Lambda_5
w_{min} \lesssim m_1^V$, we have
\begin{equation}
\label{eq:25}
\Lambda_5^2 w_{\mathrm{min}}^2 \lesssim \frac{\int
  dz\, w (\partial_z a)^2}{\int dz \, w(a-1)^2} \; .
\end{equation}
Integrating by parts and using Eq.~(\ref{eq:9})
at ${p^2 =0}$, one shows that the numerator in Eq.~(\ref{eq:25})
is smaller than $-\partial_z a|_{z_{\mathrm{UV}}}$. The denominator 
equals $\int ({a-1})^2 da \, \left(w^2/ wa'\right) \geq
-w_{\mathrm{min}}^2/3\partial_z a|_{z_{\mathrm{UV}}}$, where we
minimized the term in the parenthesis and then evaluated the
integral. One obtains 
$\Lambda_5^2w_{\mathrm{min}}^4 <  3(\partial_z a)^2_{z_{UV}} = 3[96
  \pi^2 m_W^2 / 5\Lambda_5 g^2]^2$, where Eqs.~(\ref{eq:13}),
(\ref{eq:19}) were used to express $\partial_z a|_{z_{UV}}$ and
$g_5$. We get finally ${\Lambda_5 w_{\mathrm{min}} < 6\pi m_W
/g}$ which proves the statement.

To summarize, the inequality~(\ref{eq:19}) should be valid, otherwise
the theory is no better than a 4D theory of massive $W$-bosons  without the Higgs mechanism.

%%%%%%%%%%%%%%%%%%e%%%%%%%%%%%%%%%%%%%%%%%%%%%%%%%%%%%%%%%%%%%%%%%%%%%
\section{Constraint on the $S$ parameter}
\label{sec:constr-s-param}
At the culmination of this Letter we derive a bound on $S$
  parameter from the weak coupling condition  (\ref{eq:19}). 
First, we show that the $S$ parameter, Eq.~(\ref{eq:26}), is minimal
at $v(z)=0$. To this end we find the variation $\delta
a^2(z)$ due to $\delta v^2(z)>0$ by varying and solving
Eq.~(\ref{eq:9}) at $p^2=0$,
\begin{multline}
\label{eq:29}
a(z)\delta a(z) = -2g_5^2\int dz'\, a(z)G_{p=0}^A(z,z')a(z') \\ \times
w^3(z') \,\delta v^2(z') < 0\;,
\end{multline}
where the integrand is positive in virtue of Eq.~(\ref{eq:15}). Thus, $a^2$
decreases and $S$ grows  as $v^2$ increases.

At $v=0$ we explicitly find ${a(z) = 1
- I(z)/I(z_{\mathrm{IR}})}$ by solving Eq.~(\ref{eq:9}) at
$p^2=0$. Substituting this into Eq.~(\ref{eq:26}), we get
\begin{equation}
\label{eq:23}
S > \frac{8\pi}{g_5^2}\int dz\, \frac{wI}{I(z_{\mathrm{IR}})} \geq
\frac{8\pi}{g_5^2 I(z_{IR})m_1^V} \left[\int dz\, w I\right]^{1/2}
\end{equation}
where we took into account $I(z) < I(z_{\mathrm{IR}})$ in the 
first inequality and Eq.~(\ref{eq:14}) in the second. The integral in
brackets is equal to $\int I dI\, w^2 \geq w_{\mathrm{min}}^2
I^2(z_{\mathrm{IR}})/2$. Using
Eq.~(\ref{eq:19}), we obtain
\begin{equation}
\label{eq:28}
S> \frac{8\pi w_{\mathrm{min}}} {g_5^2 m_1^V \sqrt{2}} =
\frac{5}{6\pi \sqrt{2}}\cdot  \frac{\Lambda_5 w_{\mathrm{min}}}{m_1^V}
\gg 0.2\;,
\end{equation}
in obvious conflict with the experimental data.

%%%%%%%%%%%%%%%%%%%%%%%%%%%%%%%%%%%%%%%%%%%%%%%%%%%%%%%%%%%%%%%%%%%%%%%
\section{Higher-order operators}
\label{sec:high-order-oper}
The model (\ref{eq:1}) 
is defined modulo higher-order terms in the Lagrangian
suppressed by the cutoff
$\Lambda_5$. One asks whether they can alleviate our bound on
$S$, given the consistency requirements of
Sec.~\ref{sec:model}: $v^2 \ll \Lambda_5^3$, and $w(z)$, $v(z)$ 
are nearly constant on the length scale $w(z)\Delta z \sim
\Lambda_5^{-1}$.
To this end, let us consider explicitly
the lowest of these
terms~\cite{agashe},
\begin{equation}
\label{eq:11}
\Delta {\cal S} = - \frac{c }{\Lambda_5^3} \,\int dz\, d^4 x\,
\frac{w}{2g_5^2} \, 
\mathrm{tr} \left[ L_{MN} X R_{MN} X^{\dag} 
  \right]  \;,
\end{equation}
where $c\lesssim 1$; 
parity-even terms of the same order 
are irrelevant as they can be absorbed at the quadratic level
into redefinition of $w(z)$ and~$v^2(z)$.
The correction \eqref{eq:11} changes Eqs.~(\ref{eq:16}); in
particular, Eq.~(\ref{eq:9}) becomes
\begin{equation}
\label{eq:22}
-\frac{1}{\tilde{w}}\partial_z \left(\tilde{w} \partial_z A_\mu
\right) - (p^2 - 2g_5^2\tilde{w}^2 \tilde{v}^2) A_\mu = 0\;,
\end{equation}
where $\tilde{w} = w\,  (1 + c v^2/2\Lambda_5^3)$ and $\tilde{v}^2 = v^2
w^3 / \tilde{w}^3$ contain small corrections. One calculates~$S$ and obtains, 
$$
S = \frac{8\pi}{g_5^2} \int dz\,  \tilde{w} (1-\tilde{a}^2)
-\frac{5 c}{6\pi\Lambda_5^2} \int dz \, wv^2
 = S_+ + S_-\;,
$$
where $S_+$ and $S_-$ are the first and second integral, respectively; 
$\tilde{a}(z)$ satisfies Eq.~\eqref{eq:22} at ${p^2=0}$ with boundary
conditions $\tilde{a}(z_{\mathrm{UV}}) = 1$,
$\tilde{a}(z_{\mathrm{IR}}) = 0$. 

Off hand, the term $S_-$ could lower the value of $S$ parameter.
Let us prove, however, that $|S_-| \ll |S_+|$. 
At $v^2=0$ we have
$S_-=0$, $S_+\ne 0$. 
Let us consider the variation $\delta v^2 (z)$ 
keeping  $\delta \tilde{w}(z) = 0$. The variation $\delta \tilde{a}(z)$
is again given by Eq.~(\ref{eq:29}) with $w$ and $v$
replaced by  $\tilde{w}$ and $\tilde{v}$.
The integrand in Eq.~(\ref{eq:29}) is positive and 
stays constant on the length scale $w(z)\Delta z \sim
\Lambda_5^{-1}$. 
Therefore,
$$
- \tilde{a}(z)\delta \tilde{a}(z) \gg 2g_5^2 \Delta z  \,
\tilde{a}^4(z) \tilde{I}_A(z) \tilde{w}^3(z) \,\delta
\tilde{v}^2(z)\;,
$$
where we used Eq.~(\ref{eq:15}) for the Green's function. This gives
\begin{equation}
\label{eq:31}
\delta S_+ \gg \frac{32\pi}{\Lambda_5^2}\int dz\, \tilde{w}(z)
\tilde{a}^2(z) \,\delta v^2(z)\;,
\end{equation}
where we substituted 
$\tilde{I}_A(z) \gg \Delta z / [\tilde{w}(z)\tilde{a}^2(z)]$ and ignored
small multiplicative correction to $w(z)$. We see that as $v^2$ increases,
$v^2 \to (1+\epsilon)v^2$, the term $S_+$ grows faster than $|S_-|$ unless
\begin{equation}
\label{eq:32}
\int dz \, \tilde{w} \tilde{a}^2 v^2 \ll \int dz\, w v^2\;,
\end{equation}
On the other hand, if
the inequality \eqref{eq:32} is satisfied, then
\begin{equation}
\notag
S_+ > \frac{8\pi}{g_5^2} \int dz \, \tilde{w} \,\frac{v^2}{v_{\max}^2} \,
\left[1-\tilde{a}^2 \right] \approx \frac{\Lambda_5^3}{ c
  v_{\max}^2} |S_-| \gg |S_-|\;,
\end{equation}
where we inserted $v^2/v_{\max}^2 < 1$ in the integrand, ignored the
second term in brackets due to Eq.~(\ref{eq:32}) and expressed the
result in terms of $S_-$. 

So, we proved that $S \approx S_+$.
The analysis of Sec.~\ref{sec:constr-s-param} goes through for $S_+$
and yields the bound (\ref{eq:28}).
In this way we come to the intuitively clear conclusion that the
higher-order term \eqref{eq:11} does not affect this bound 
  (see Ref.~\cite{agashe} for the limited numerical analysis of the
same problem). Operators of even higher orders should
be even less important. 

%%%%%%%%%%%%%%%%%%%%%%%%%%%%%%%%%%%%%%%%%%%%%%%%%%%%%%%%%%%%%%%%%%%%%%%
\section{Conclusions}
\label{sec:conclusions-1}
In this Letter we derived the weak coupling condition ${m_1^V \ll
  \Lambda_5 w_{\mathrm{min}}}$, where $m_1^V$ is the lowest
Kaluza-Klein mass, $\Lambda_5 w_{\mathrm{min}}$ is the redshifted
cutoff. We proved that within the holographic technicolor models
defined by Eqs.~(\ref{eq:1}),~(\ref{eq:4}), this condition
bounds  the value of $S$ parameter, $S\gg 0.2$, in conflict with
experimental data. The latter bound is stable
with respect to higher-order corrections and agrees with the 
conjectured general 
constraint of Refs.~\cite{Sannino}.

 One can interpret our results in terms of 4D technicolor theory by
  recalling that $S\propto 1/g_5^2 \propto N_c$, where $N_c$ is the
  number of technicolors. Thus, $S$ is smallest at $N_c\sim 1$ when
  holography is not trustworthy. The opposite requirement of weakly coupled
  holographic description leads to a lower bound on $S$ which, as we
  demonstrated, is $S\gg 0.2$. 

Since the troubles come from vectors and axial vectors,
our bound can possibly be avoided in models with 
modified vector sectors. One can think of changing the boundary
conditions (\ref{eq:4})~\cite{hong, piai} or considering parity
breaking~\cite{piai}.\footnote{One can also lower $S$ by
  loop contributions from extra fermions~\cite{extra_fermions}.} 
In any case healthy models should be special, as
our results suggest. If constructed, they would shed light on the
structure of phenomenologically viable technicolor theories.

\paragraph*{Acknowledgments} We thank M.V.~Libanov,
S.M.~Sibiryakov and P.G.~Tinyakov for discussions. This work was
supported by the Dynasty foundation (S.T. and Y.Z.) and grants RFBR
12-02-01203 (D.L., S.T., and Y.Z.), RFBR 10-02-01406, RFBR 11-02-01528
(S.T.), SCOPES (V.R.), NS-5590.2012.2.

%%%%%%%%%%%%%%%%%%%%%%%%%%%%%%%%%%%%%%%%%%%%%%%%%%%%%%%%%%%%%%%%%%%%%%

\end{document}